Low-Temperature Co-Fired Ceramics for a Sustainable Planar Plasma Jet with Homogeneous Plasma in Large Treatment Areas for Biomedical Applications

*Ivan Gomez Ho[1], Hua-Lin Chen[1,3], Cheng-Han Tsai[1], Jong-Shinn Wu[1], Yun-Chien Cheng[1,2]*

[1]Department of Mechanical Engineering, National Yang Ming Chiao Tung University, Hsinchu 30010, Taiwan

[2]Department of Electrical Engineering, National Taiwan University, Taipei, Taiwan

[3]National Center for Instrumentation Research, NIAR, Hsinchu, Taiwan

Corresponding author: Yun-Chien Cheng

E-mail: yccheng@nycu.edu.tw

## Abstract

This study presents a portable planar argon-based plasma jet designed for large-area biomedical applications, with an emphasis on uniform discharge, stability, and safety. The device incorporates interchangeable rear and side inlet gas adapters with restriction plates and channels to equalize gas flow, while electrodes are encapsulated in low-temperature co-fired ceramics to reduce degradation and arcing during repeated operation. Flow simulations were conducted for multiple channel configurations to ensure laminar gas distribution and uniform plasma generation. Device performance and safety were evaluated using the kinPen MED as a reference standard. Electrical characteristics, optical emission, discharge uniformity, temperature, gas velocity, ozone generation, UV irradiance, and leakage current were systematically measured. The plasma exhibited stable voltage, current, and power after an initial warm-up period. Temperatures stabilized within minutes, with the rear-inlet configuration demonstrating lower operating temperatures due to higher outlet gas velocity. Ozone levels remained below established safety limits, while UV exposure constrained allowable treatment times. Leakage current decreased with increasing distance and approached safety thresholds at short separations. These results demonstrate that the proposed planar plasma jet provides stable, uniform plasma delivery while meeting key safety requirements, supporting its potential for clinical and biomedical use.

## 1. Introduction

Plasma, often termed the fourth state of matter, consists of ions, electrons, and reactive species, formed when sufficient energy is added to a gas to ionize it. Cold atmospheric plasma (CAP), generated at near-atmospheric pressure and close to room temperature, has attracted considerable interest in biomedical applications. Due to its composition of free radicals, and reactive oxygen and nitrogen species (RONS), CAP exhibits antimicrobial activity and has been applied in skin treatment, wound healing, and cancer therapy [1,2].

CAP can be generated using dielectric barrier discharges (DBDs) and plasma jets. in which electrodes are separated by dielectric layers. Once electrical breakdown occurs, multiple micro discharges form, ionizing the working gas and sustaining plasma generation. Plasma jets, in contrast, employ gas flow through a dielectric tube to generate a downstream plasma plume. While both approaches are widely used, each presents limitations for biomedical applications. plasma jets produce localized plumes suited for spot treatments, while planar DBDs has wide area coverage, but lacks guided gas flow and do not generate non-contact plumes [3–10].

A common issue across many CAP device designs is electrode exposure. Electrodes that are uncovered or only partially shielded by dielectric materials are prone to oxidation and contamination during repeated operations. These effects can alter discharge characteristics over time and promote non-uniform behavior such as arcing, which compromises stability and safety. To address these limitations, this study employs a planar plasma jet (PPJ) that combines wide-area coverage with guided gas flow to generate a non-contact plume, while fully encapsulating electrodes to enhance stability and durability.

PPJs are designed to deliver a broad and more uniform plasma sheet particularly beneficial for treatments requiring even plasma exposure across large areas [11, 12]. Despite their advantages multiple studies [13–20] report lateral non-uniformity in PPJ plumes, often with brighter edges and a dimmer center at the outlet. This behavior is frequently attributed to the use of rod-like electrodes positioned near the slit edges, which intensify the electric field locally and lead to edge-brightened or gliding-arc–like discharges. These findings demonstrate the necessity of full dielectric encapsulation to ensure stable, safe plasma operation.

Gas flow conditions also play a critical role in plume stability and uniformity, with low flow rates causing asymmetry and higher flow rates promoting extended, but brush-like discharges [20-22]. Gas temperature has likewise been shown to decrease with increasing flow rate, with reported values of 1,200–2,000 K for PPJ's [23]. Additionally, exposed or partially dielectric-covered electrodes in PPJs have been associated with overheating and arcing, limiting plume length and long-term sustainability [13–15,20].

In this study, Low-Temperature Co-fired Ceramic (LTCC) technology is employed to fabricate a planar plasma jet. LTCC enables multilayer device architectures in which electrodes can be hermetically sealed within ceramic layers, providing protection against oxidation and contamination [24–26]. Although LTCC has been explored for plasma devices, prior studies reported electrode overheating, thermal stress, limited operational lifetime, and complex fabrication processes such as laser ablation [27–29]. These limitations highlight the need for improved LTCC-based designs optimized for sustained plasma generation.

In summary, this study designs and evaluates an LTCC-based planar plasma jet, focusing on gas flow uniformity, electrical stability, and operational safety for biomedical applications.

# 2. Materials and Methods

## 2.1 Criteria for Sustainability and Robustness

| Criteria | Parameters |
|---|---|
| Plasma Discharge Time Duration | 1-2 hours |
| Temperature of LTCC Device | Below 50 °C |
| Gas Temperature | Below 40 °C |
| Ozone concentration | 0.10–0.13 ppm |
| UV concentration | 30 J/m$^2$ |
| Current | ≤100μA |

Table 1. Experimental measurement criteria

Table 1 summarizes the evaluation criteria used in this study, based on the kinPen MED [30] and ICNIRP safety guidelines [31].

## 2.2 LTCC PPJ Equipment and Experimental Setup

The diagram of the experimental setup is shown in Figure 1. The LTCC PPJ was connected to high-voltage, ground, and gas supply lines. The power supply (HVPS-S10, Naiding Technology, Taiwan) used in this study provided an AC voltage of 6.2kV with a frequency of 21.55kHz. Voltage and current were measured using a high-voltage probe (P6015A, Tektronix, United States) and a Rogowski coil (Current Monitor Model CM-100-MG, Ion physics corporation, United States) with waveforms recorded on an oscilloscope (WaveSurfer 3054z, Teledyne Technologies, United States). Argon gas was used as the working gas at a controlled flow rate of 20 standard liters per minute (SLM) through a flow meter (0-20 SLM, Yongxin Instrument, Taiwan).

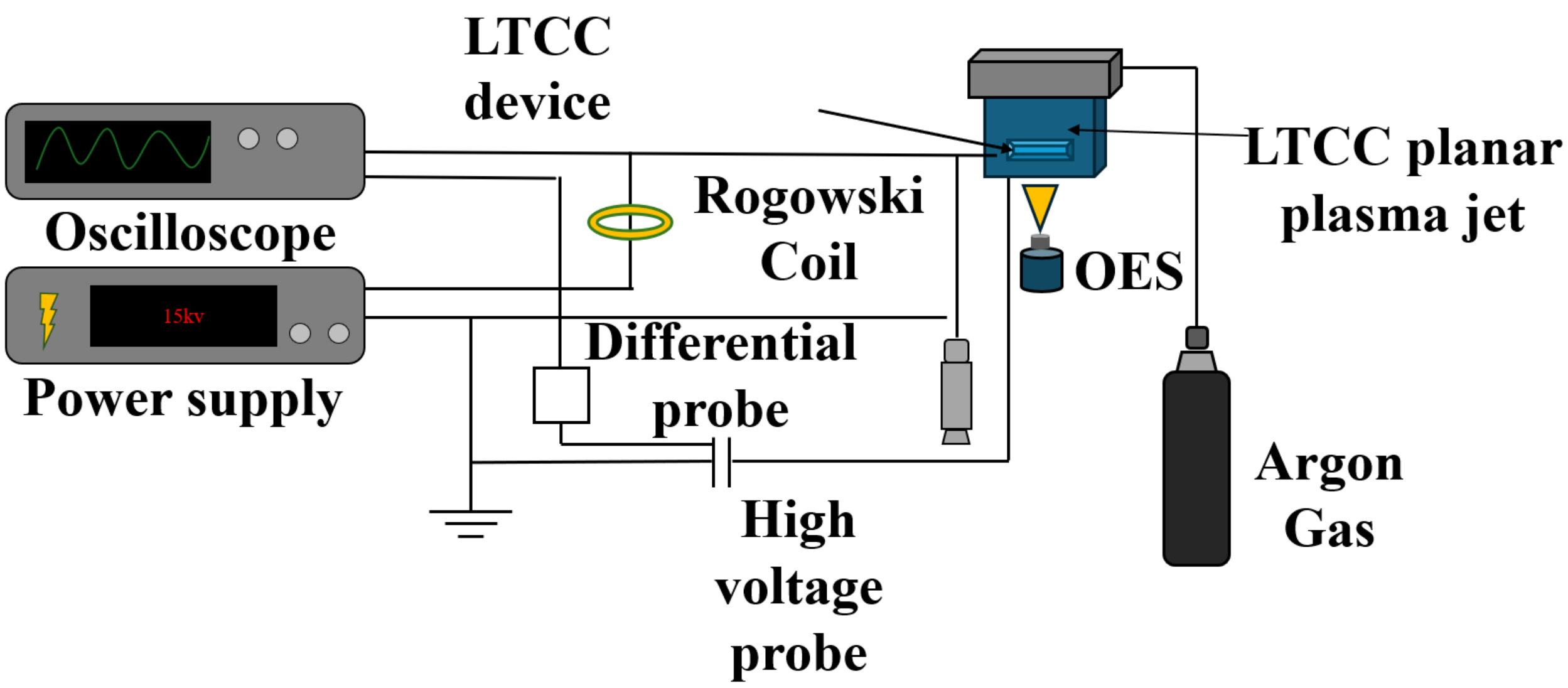

Figure 1. Experimental setup diagram

### 2.3 LTCC Plasma Generator and Gas Adapter Designs

The plasma device employed in this study uses low-temperature co-fired ceramic (LTCC) as the dielectric. The electrodes are hermetically encapsulated within the LTCC to isolate them from ambient air. For large-area coverage, electrodes were patterned as long, narrow strips (6 × 29 mm), while the LTCC body measured (40 × 20 × 0.9 mm). This design is based upon previous work developed for large-area wound treatment [32].

### 2.4 LTCC Gas Adapter Design

Two interchangeable gas adapters, a rear-inlet and a side-inlet design, were developed for the LTCC-based PPJ (Figure. 2), differing primarily in their gas flow characteristics. The rear-inlet design increases gas velocity to extend plume length, whereas the side-inlet design promotes more uniform gas distribution by allowing gas to first mix within a mixing chamber before flowing out.

Each adapter includes an upstream restriction plate to improve flow uniformity. The plate contains 11 cavities: 2 mm diameter holes at the center and ends to enhance gas delivery to critical regions, and 1.5 mm diameter holes for the remainders. Hole spacing was 3 mm, increasing to 3.5 mm at the center, to promote uniform gas distribution.

Both adapters were 3D printed with 6, 7, or 9 internal channels to enhance gas velocity and flow uniformity. Channel length was fixed at 3 mm, while divider widths were adjusted to 3, 2, and 0.75 mm, respectively, based on channel number.

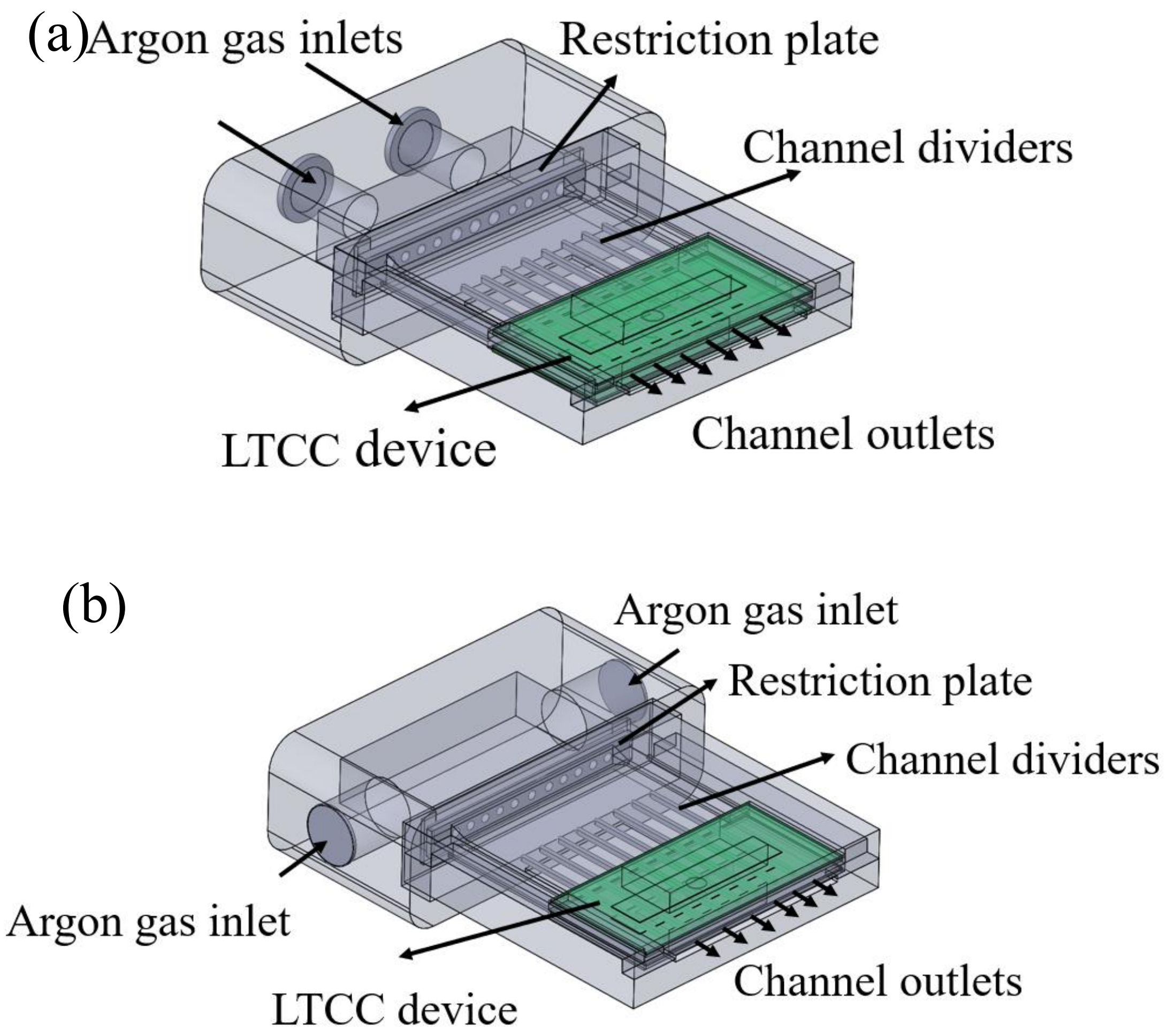

Figure 2. (a) Rear inlet gas adapter, (b) Side inlet gas adapter

## 2.5 CAP Measurement Instruments

### 2.5.1 Optical Emission Spectroscopy

Optical emission spectroscopy (OES; Acton SP2500) was used to evaluate plasma uniformity and UV emission. The OES lens was positioned 4 cm away from the plasma discharge and aligned at 0 mm height from the outlet of the LTCC device. The intensity of argon plasma was then measured at wavelengths ranging from 180 nm to 900 nm (Figure 1).

### 2.5.2 Lissajous Figures

Lissajous figures and total power consumption were obtained by connecting the plasma device in series with a 330-pF metal film capacitor (WIMA FKP3 330P). The voltage was measured using a differential probe (SI-9002, Tiepie, Netherlands) (Figure 1). Power consumption was calculated from the Lissajous figures using MATLAB (Matlab, The MathWorks, United States).

### 2.5.3 Temperature measurement by Forward Looking Infrared (FLIR)

An infrared thermal imaging camera (FLIR ONE® Edge Pro, Teledyne Technologies Incorporated, USA) was used to record the temperature over a duration of 30 minutes. Temperature measurements were taken from the surface of the LTCC device during operation. Additionally, gas temperature was measured by discharging plasma onto porcine skin for 5 minutes and then measuring its temperature.

### 2.5.4 Ozone Gas Analyzer

An ozone gas analyzer (EST-1015H, Environmental Sensor Technology, USA) was used to measure ozone concentration produced by the plasma discharge throughout its operation. The measurable concentration range is between 0-5 ppm with a resolution of 0.01ppm and an accuracy of ±5% FS.

### 2.5.5 Intensified Charge-Coupled Device (ICCD) Camera

An ICCD camera (PI-MAX4 1024i, Teledyne Princeton Instruments, USA) was used to capture time-gated, spatially resolved emission images of the PPJ, mapping plasma light intensity across individual channels within the device.

### 2.5.6 Patient Leakage Current Measurement

Patient leakage current was measured using a human body model based on prior work [33]. As shown in Figure 3 an aluminum tape electrode was connected to a human body model circuit consisting of resistors R1 (10 kΩ), R2 (1 kΩ), and a capacitor C1 (0.015 µF). Voltage across the capacitor was measured using a differential probe and used to calculate leakage current.

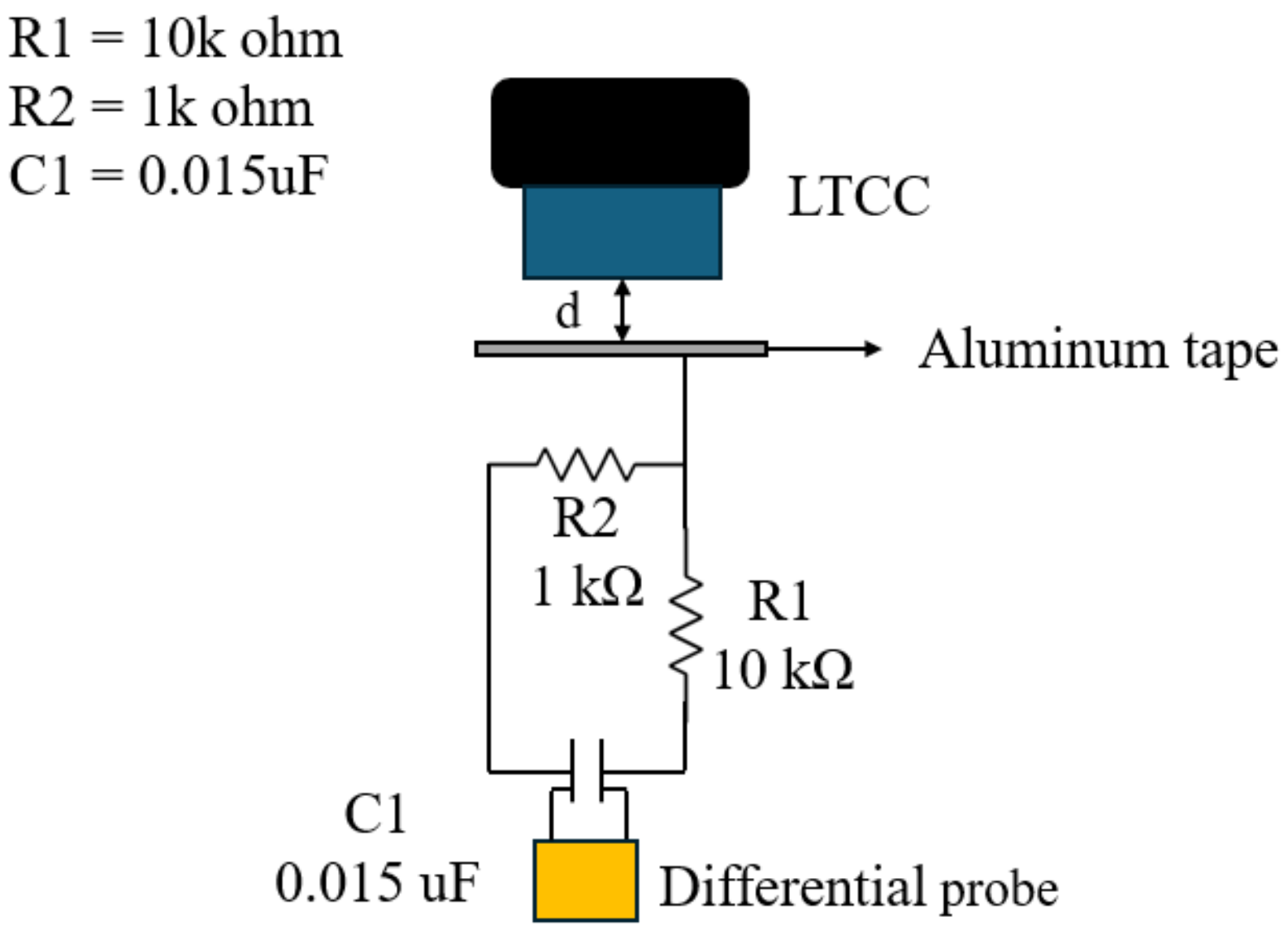

Figure 3. Electrical human body model diagram

#### 2.5.7 Gas Flow Velocity

Gas flow velocity at the PPJ outlet was measured using an anemometer (GM816, BENETECH, Shenzhen) placed 2 cm from the device. The range for measurable air velocity is 0-30 m/s with a resolution of 0.1 and an accuracy of ±5%.

## 3. Results

### 3.1 Gas Flow Simulations

#### 3.1.1 Gas Flow Velocity and Uniformity Simulations

Figure 4 presents simulated gas velocity and distribution for the rear- and side-inlet adapters. The 6-channel configurations produced the highest gas velocities, whereas the 9-channel configuration exhibited the most uniform outlet distribution.

As summarized in Figure 4(g) and Table 2, velocity differences between the 7- and 9-channel configurations were small, while lower velocities corresponded to reduced deviation of velocity between channels, indicating improved uniformity. Rear-inlet designs consistently produce higher velocities than side-inlet designs. Despite having small differences (7 and 9 channel), P-value analysis (Table 2) confirms statistically significant differences among configurations.

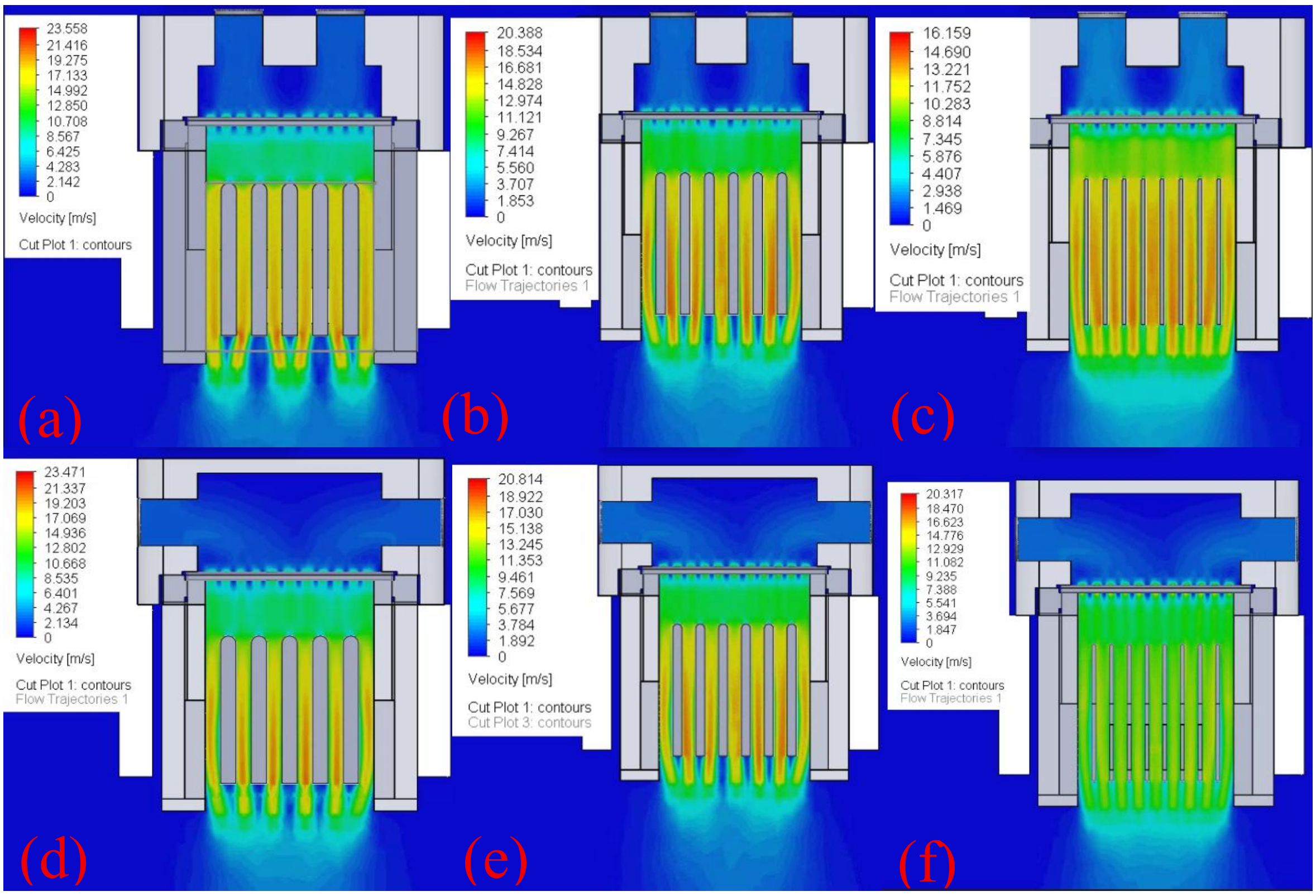

Figure 4. Simulation of gas flow velocity and distribution (a) rear inlet 6 channels (b) rear inlet 7 channels (c) rear inlet 9 channels (d) side inlet 6 channels (e) side inlet 7 channels (f) side inlet 9 channels (g) gas flow velocity plot for different inlet setups

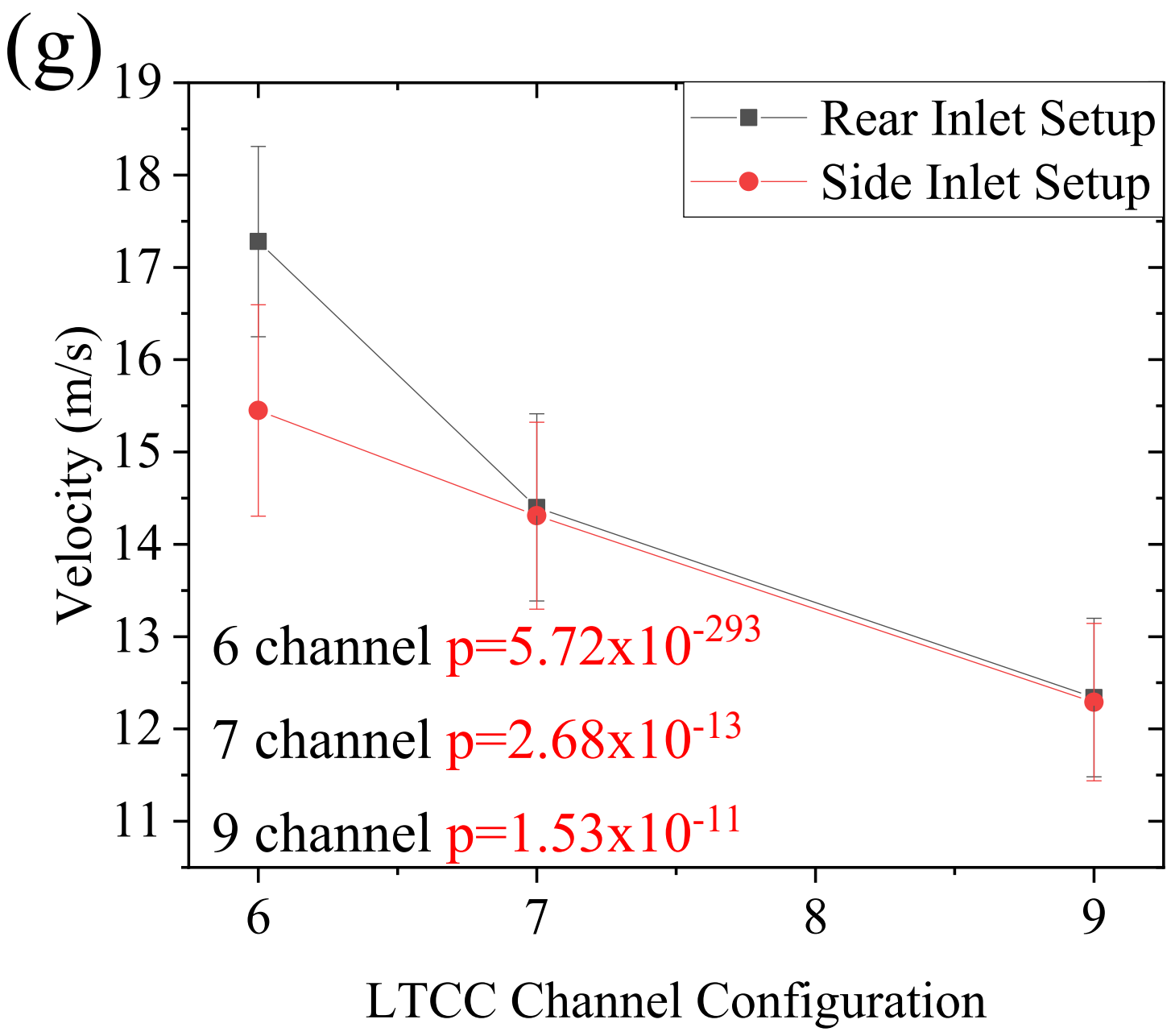

| LTCC Configuration | Velocity (m/s) | Reynolds number | P-Value |
|---|---|---|---|
| 6 Channel Rear Inlet | 17.28 | 2755 | $5.72x10^{-293}$ |
| 6 Channel Side Inlet | 15.45 | 2462 | |
| 7 Channel Rear Inlet | 14.40 | 2295 | $2.68x10^{-13}$ |
| 7 Channel Side Inlet | 14.31 | 2280 | |
| 9 Channel Rear Inlet | 12.34 | 1967 | $1.53x10^{-11}$ |
| 9 Channel Side Inlet | 12.29 | 1959 | |

Table 2. Gas adapter configurations velocity, Reynolds number and P-value

Reynolds number analysis (Table 3) shows that 6-channel configurations operate in the transitional regime, 7-channel configurations lie near the laminar–transitional threshold, and 9-channel configurations remain fully laminar ($Re < 2300$) due to lower velocities. This is due to their lower gas flow velocities. Based on these results, the 9-channel configuration is the most optimal among the configurations tested, as it ensures stable laminar flow conditions.

### 3.1.2 Air and Argon Volumes Simulation

Figure 5(g) and Table 4 show that 6- and 7-channel configurations exhibit greater ambient air diffusion than the 9-channel configuration, due to larger channel gaps. Table 4 further confirms this, showing that the 9-channel configurations for both rear and side inlets have a significantly lower air volume of just 0.24%. Air volume fractions were lowest for the 9-channel rear-inlet configuration ($2.37 \times 10^{-3}$), indicating improved argon retention.

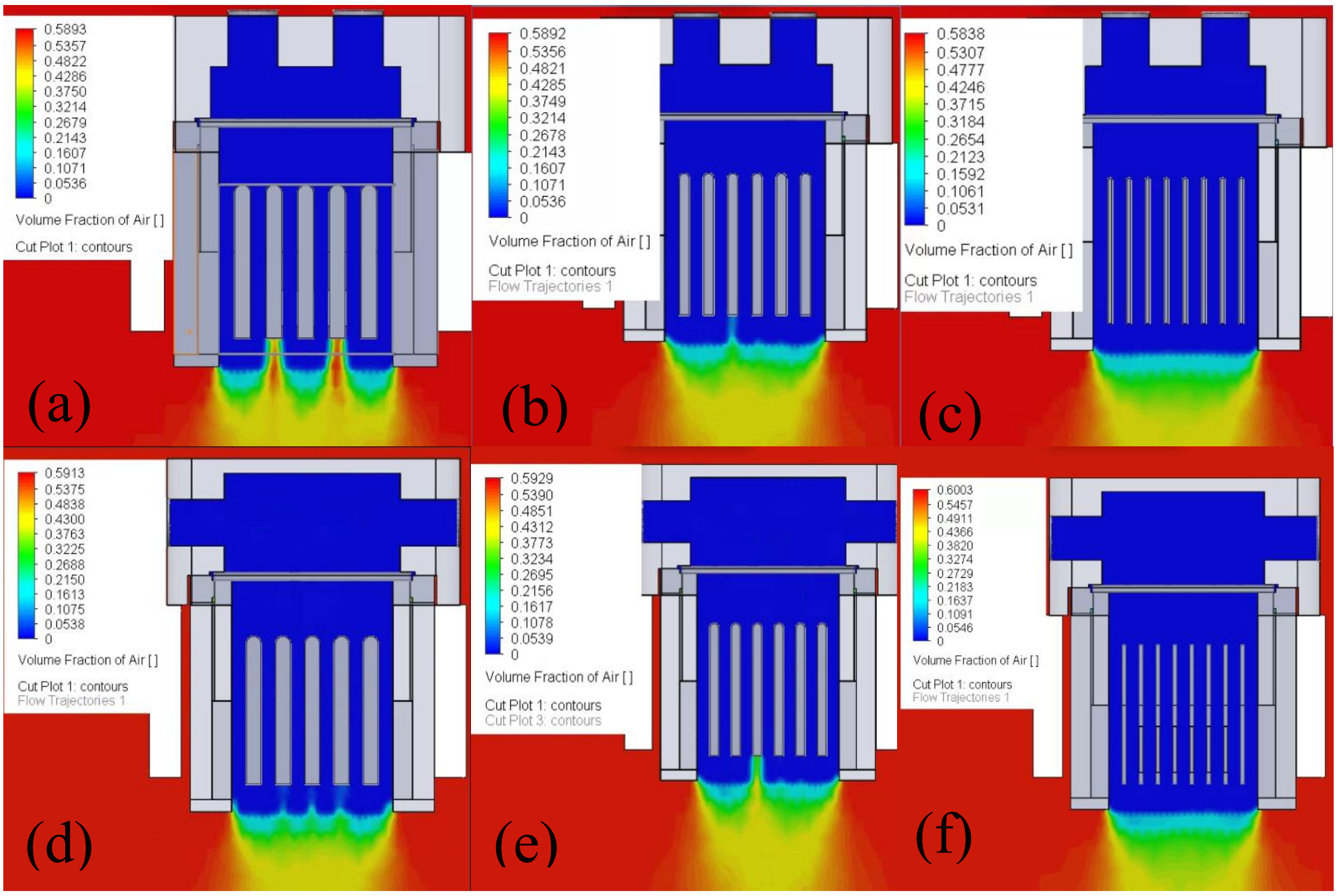

Figure 5. Simulation of air volume diffusion (a) rear inlet 6 channels (b) rear inlet 7 channels (c) rear inlet 9 channels (d) side inlet 6 channels (e) side inlet 7 channels (f) side inlet 9 channels (g) air volume plot for different inlet setups

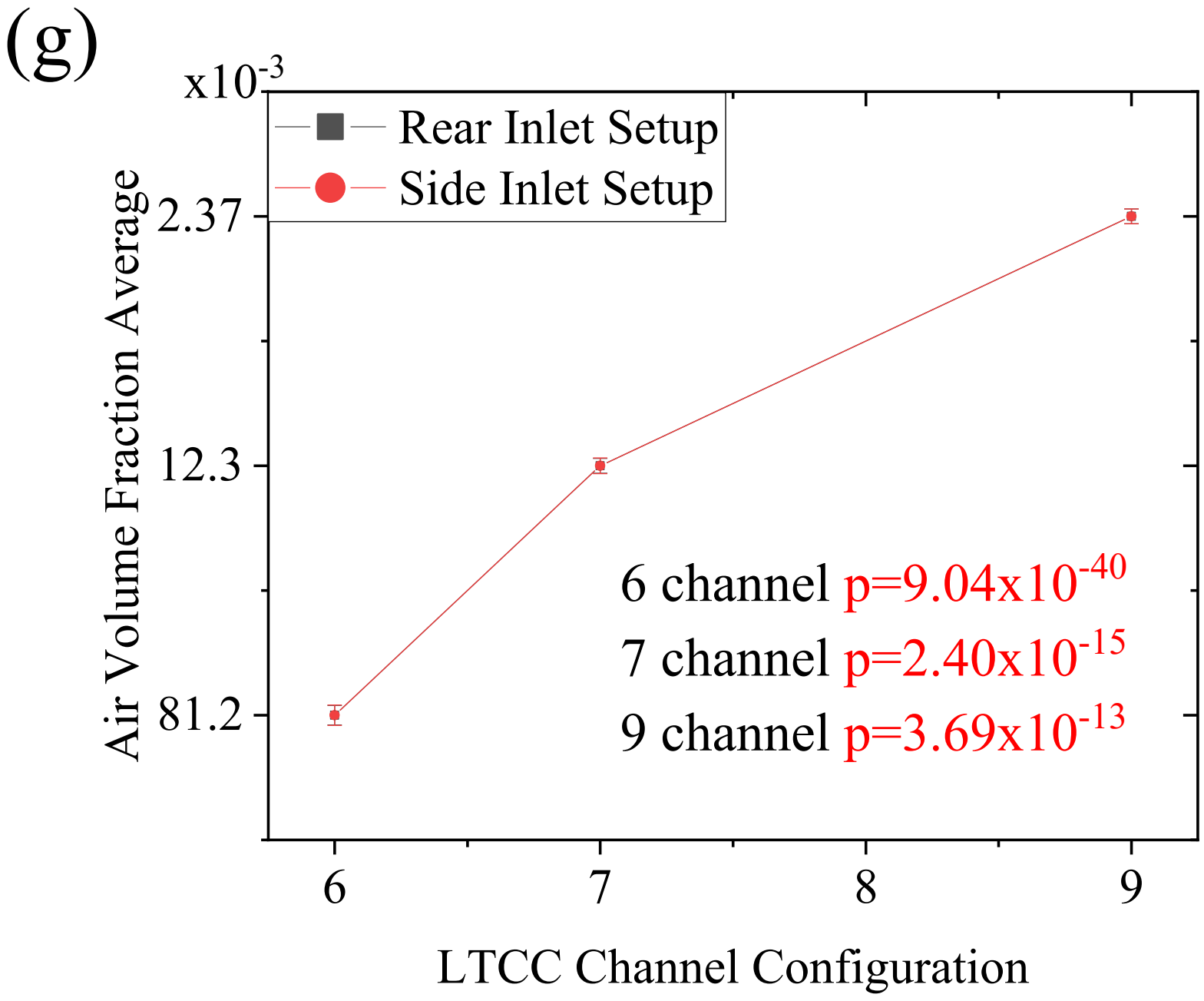

| LTCC Configuration | Air Volume Fraction Average | Volume Percentage | P-Value |
|---|---|---|---|
| 6 Channel Rear Inlet | $81.2 \times 10^{-3}$ | 8.12% | $9.04 \times 10^{-40}$ |
| 6 Channel Side Inlet | $63.4 \times 10^{-3}$ | 6.34% | |
| 7 Channel Rear Inlet | $10.4 \times 10^{-3}$ | 1.04% | $2.40 \times 10^{-15}$ |
| 7 Channel Side Inlet | $12.7 \times 10^{-3}$ | 1.27% | |
| 9 Channel Rear Inlet | $2.37 \times 10^{-3}$ | 0.24% | $3.69 \times 10^{-13}$ |
| 9 Channel Side Inlet | $2.41 \times 10^{-3}$ | 0.24% | |

Table 3. Air volume fraction measurements

Argon-volume simulations (Figure. 6, Table 5) show that 9-channel configurations retain the highest argon fraction (99.76%). Although both inlet designs achieve similar values, the rear-inlet configuration exhibits lower air intrusion (p-value<0.005), making it optimal for subsequent experiments.

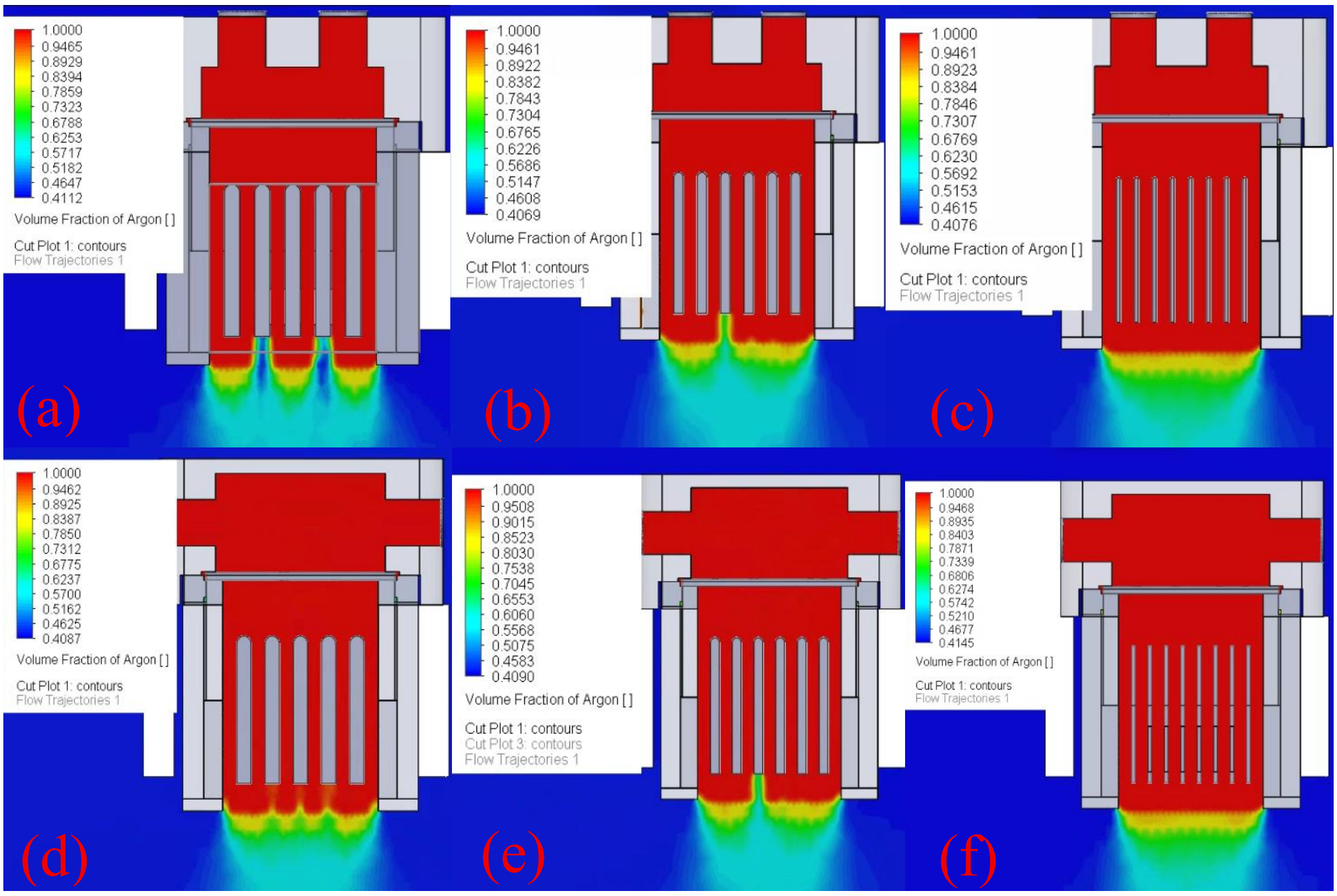

Figure 6. Simulation of Ar volume diffusion (a) rear inlet 6 channels (b) rear inlet 7 channels (c) rear inlet 9 channels (d) side inlet 6 channels (e) side inlet 7 channels (f) side inlet 9 channels (g) argon volume plot for different inlet setups

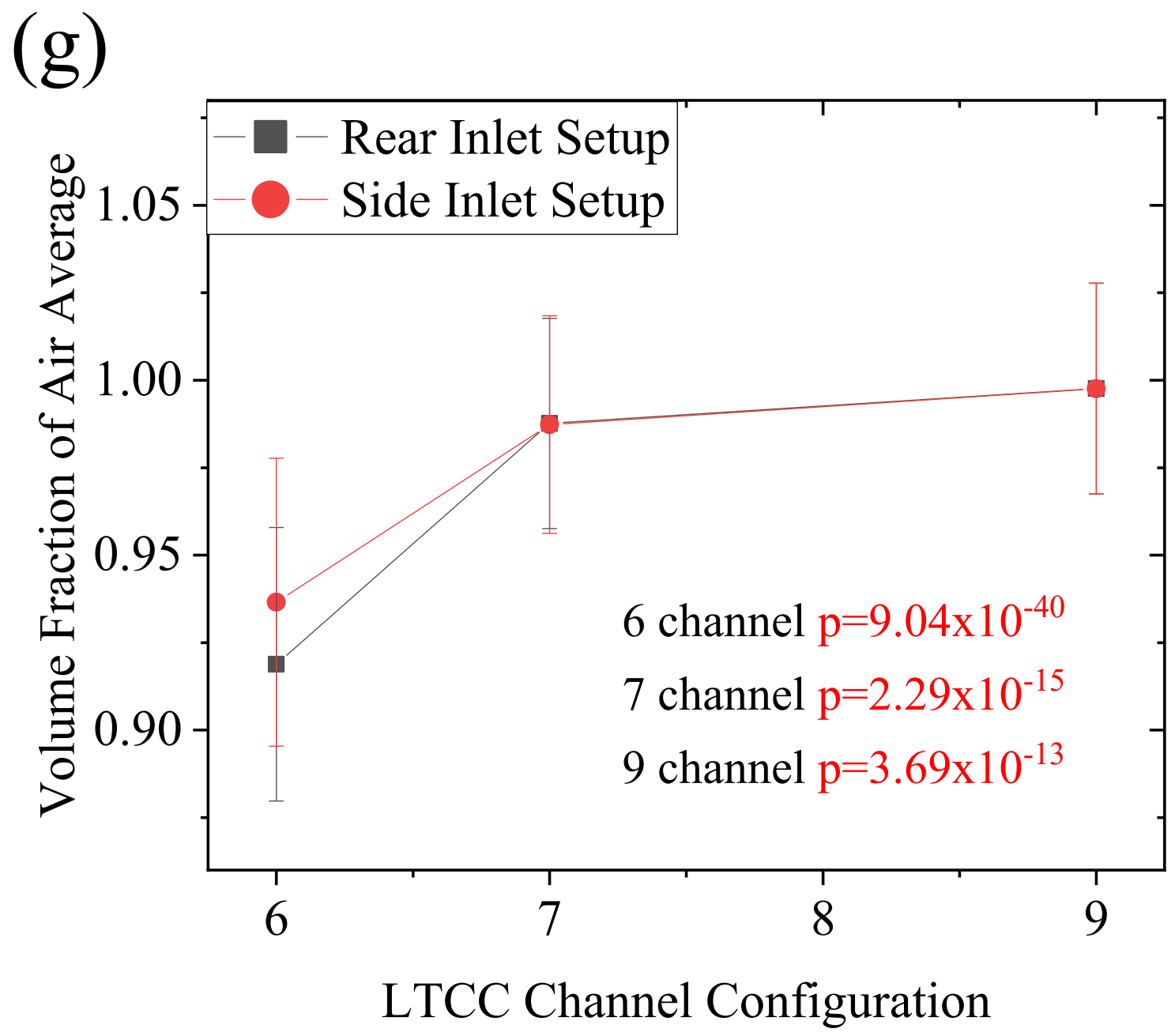

| LTCC Configuration | Argon Volume Fraction Average | Volume Percentage | P-Value |
|---|---|---|---|
| 6 Channel Rear Inlet | $91.88 \times 10^{-2}$ | 91.88% | $9.04 \times 10^{-40}$ |
| 6 Channel Side Inlet | $93.66 \times 10^{-2}$ | 93.66% | |
| 7 Channel Rear Inlet | $98.96 \times 10^{-2}$ | 98.96% | $2.29 \times 10^{-15}$ |
| 7 Channel Side Inlet | $98.73 \times 10^{-2}$ | 98.73% | |
| 9 Channel Rear Inlet | $99.76 \times 10^{-2}$ | 99.76% | $3.69 \times 10^{-13}$ |
| 9 Channel Side Inlet | $99.76 \times 10^{-2}$ | 99.76% | |

Table 4. Argon volume fraction measurements

### 3.1.3 Pressure Difference Simulation

Figure 7 presents simulated pressure differences between the device interior and ambient air. The results indicate that pressure differences significantly impact the velocity of the gas flow. Lower pressures were observed in 6-channel configurations due to higher velocities, consistent with Bernoulli's principle. This difference is due to the number of channels, where increasing the number of channels increases the cross-sectional area, leading to higher pressure. Table 6 shows that 9-channel configurations maintain higher internal pressure, therefore reducing ambient-air diffusion and improving flow stability.

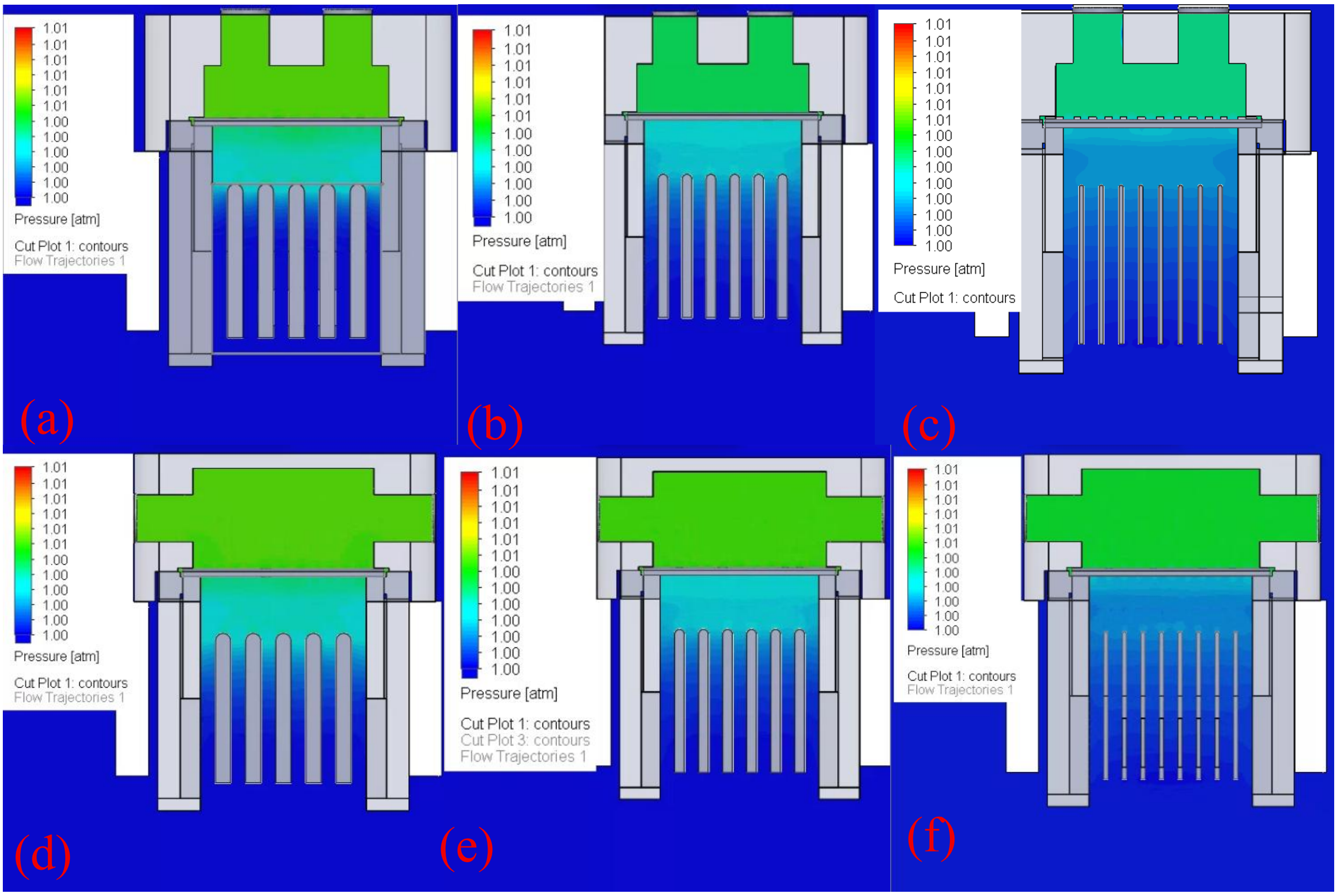

Figure 7. Simulation of LTCC system pressure (a) rear inlet 6 channels (b) rear inlet 7 channels (c) rear inlet 9 channels (d) side inlet 6 channels (e) side inlet 7 channels (f) side inlet 9 channels (g) pressure plot for different inlet setups

(g)

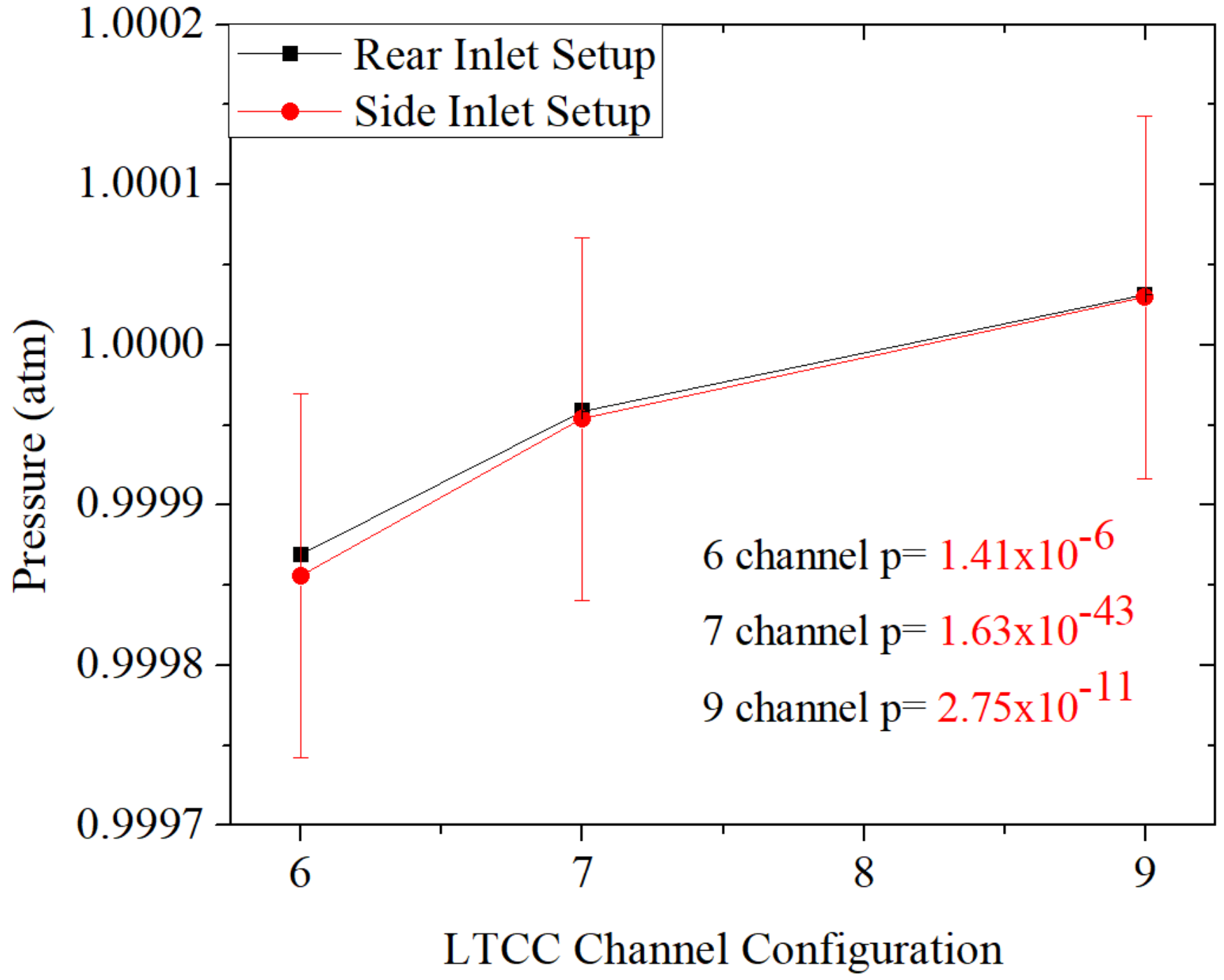

| LTCC Configuration | Pressure (atm) | P-Value |
|---|---|---|
| 6 Channel Rear Inlet | 0.99986 | $1.41 \times 10^{-6}$ |
| 6 Channel Side Inlet | 0.99987 | |
| 7 Channel Rear Inlet | 0.99995 | $1.63 \times 10^{-43}$ |
| 7 Channel Side Inlet | 0.99996 | |
| 9 Channel Rear Inlet | 1.0000 | $2.75 \times 10^{-11}$ |
| 9 Channel Side Inlet | 1.0000 | |

Table 5. LTCC system pressure measurements

Overall, the nine-channel configuration provides the most uniform flow with minimal air entrainment for both inlet designs. Although the seven-channel configuration remains laminar, increased air diffusion is expected to degrade plume uniformity in practice. Therefore, the nine-channel design was selected for fabrication and experimental validation.

### 3.2 Plasma Discharge Uniformity

#### 3.2.1 Plume Discharge Analysis

Figure 8 presents the downstream plume produced by both PPJ setups. Although plume formation requires 14–15 kV and results in filamentary discharges exceeding biomedical limits, all experiments were conducted at 6.2 kV. At this voltage, analysis was focused on internal discharge uniformity while maintaining inlet-configuration comparisons.

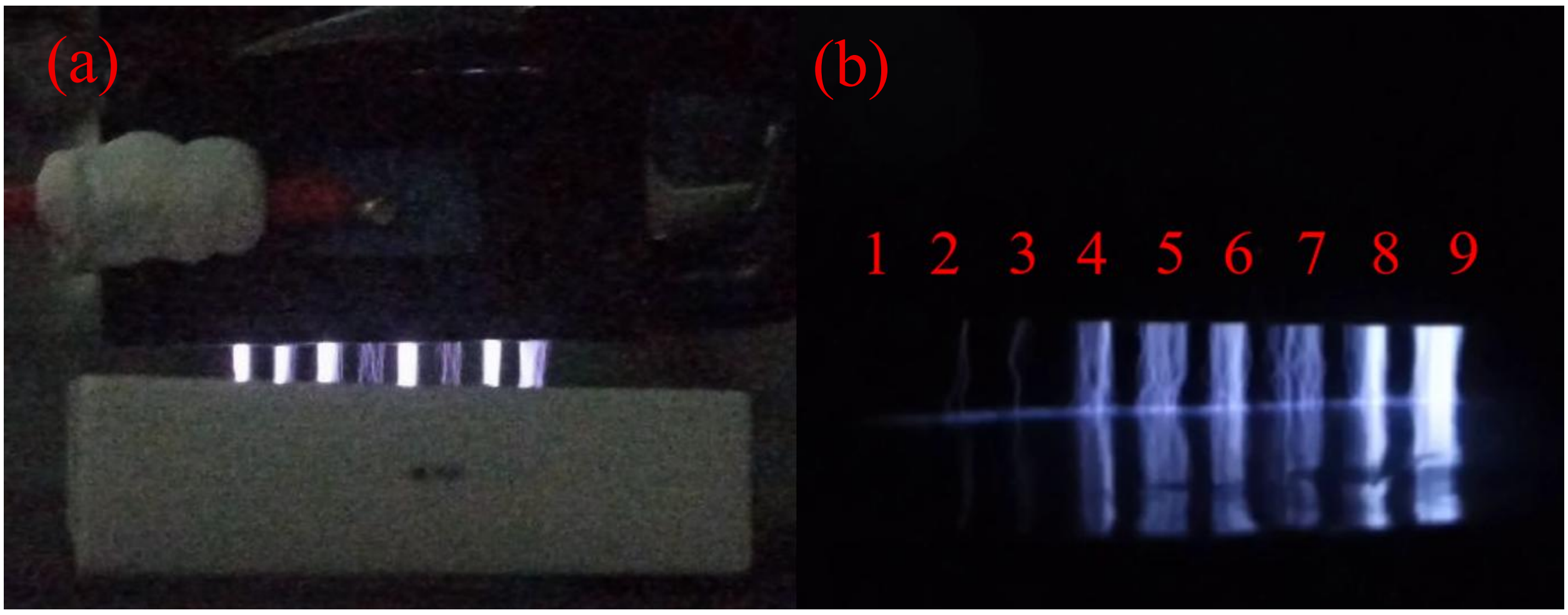

Figure 8. LTCC PPJ plasma discharge, (a) rear inlet setup, (b) side inlet setup

#### 3.2.2 Gas Flow Velocity Analysis

Outlet gas velocities for the 9-channel configuration were measured across three sections due to the narrow-slit geometry (Figure. 9). Rear-inlet designs exhibited higher velocities across all channels, consistent with simulations, with statistically significant differences ($P = 0.028$). Additionally, both configurations display variation between channels, indicating a non-uniform flow distribution along the slit.

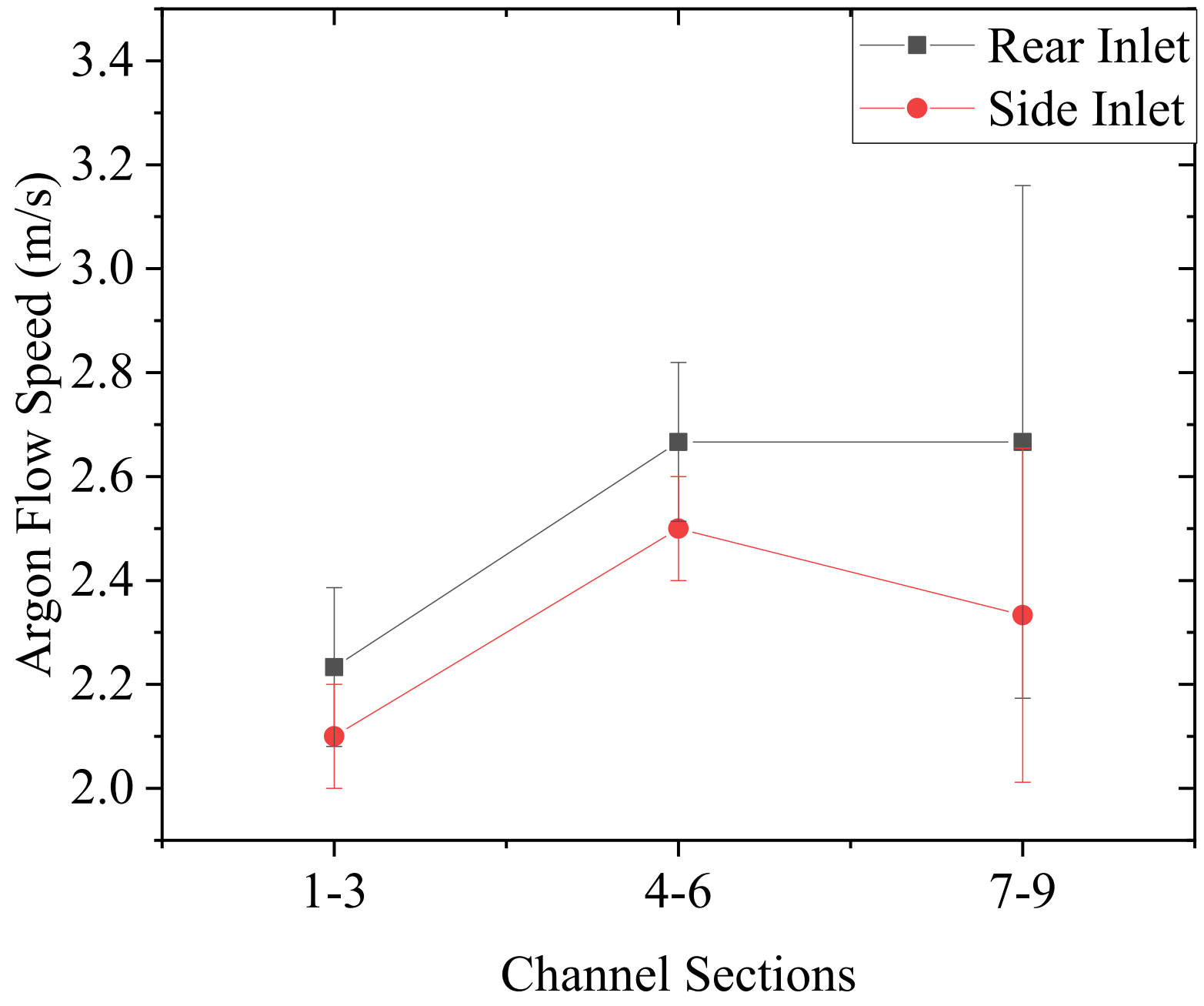

Figure 9. Argon gas flow velocity measurement

### 3.2.3 Analysis of Argon Emission Intensity

For spectral analysis, Argon emission was measured channel-by-channel at the Ar I line near 763 nm, the strongest spectral peak (Figure. 10(a)). Both inlet configurations exhibited the lowest emission at channels 1, 5, and 9, indicating a non-uniform gas distribution across these channels. In particular, the edge deficits at channels 1 and 9 are consistent with the outlet-velocity results previously shown, which demonstrated reduced speeds in channels 1–3 and 7–9. The side-inlet configuration exhibited higher emission intensity across all channels ($P = 4.15 \times 10^{-4}$). This result aligns with the velocity data, higher rear-inlet velocities reduce gas residence time, decreasing ionization probability and detected emission intensity.

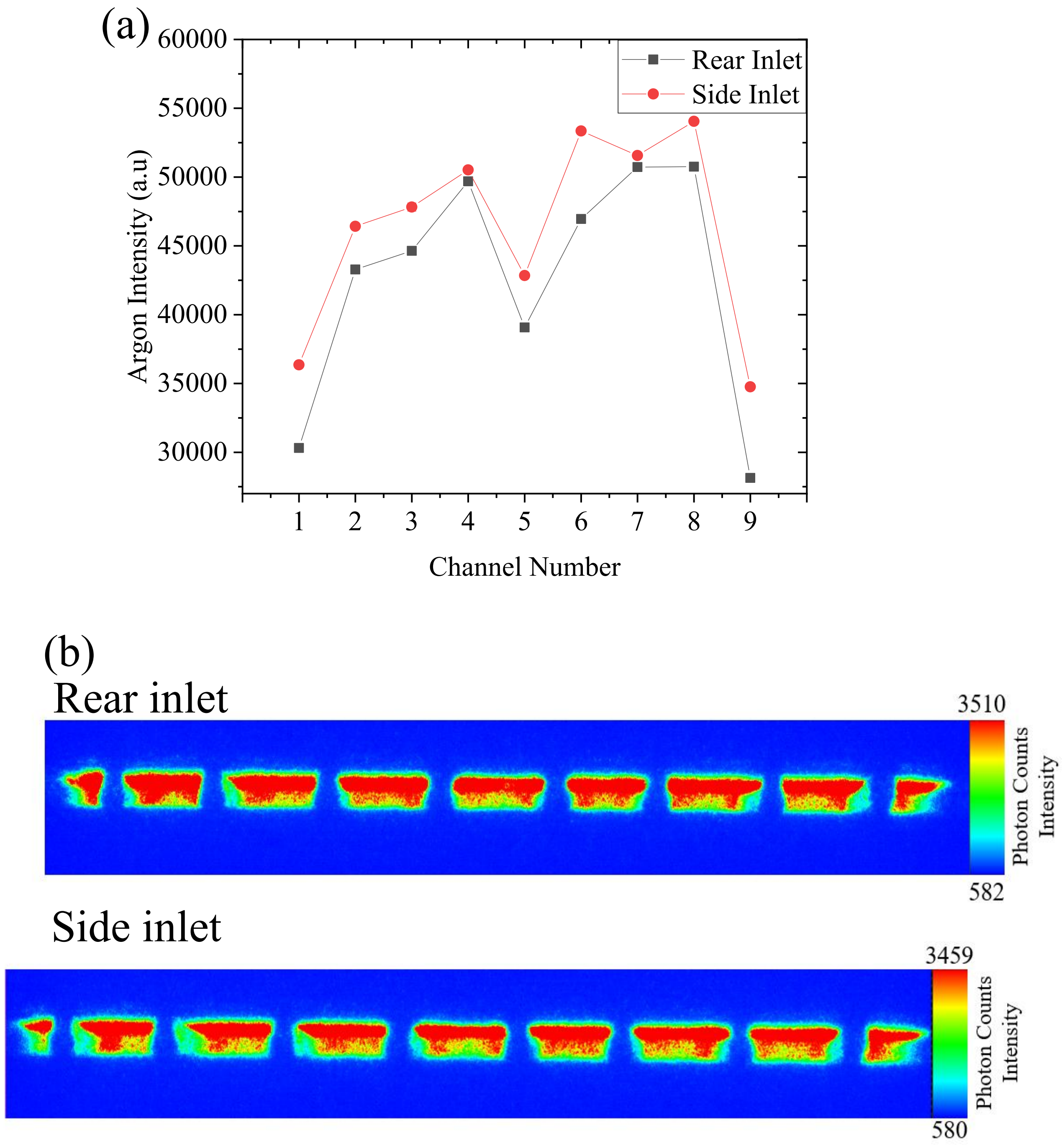

Figure 10. (a) OES analysis for Argon emission for rear and side inlet configuration, (b) ICCD imaging for LTCC PPJ rear and side inlets

### 3.2.4 ICCD Plasma Intensity Imaging

To further assess PPJ uniformity, ICCD images of the discharge were acquired. In contrast to OES, which records specific emission lines, the ICCD was used without an optical filter in this study, capturing all light within its spectral response and reports photon counts over the gated exposure. As shown in Figure 10, the side-inlet configuration (Figure 10 (b)) exhibits higher integrated intensity than the rear-inlet, confirming the OES finding that argon emission is stronger for the side-inlet setup.

### 3.3 LTCC PPJ Electrical Stability and Safety

#### 3.3.1 LTCC Plasma Device Voltage Stability

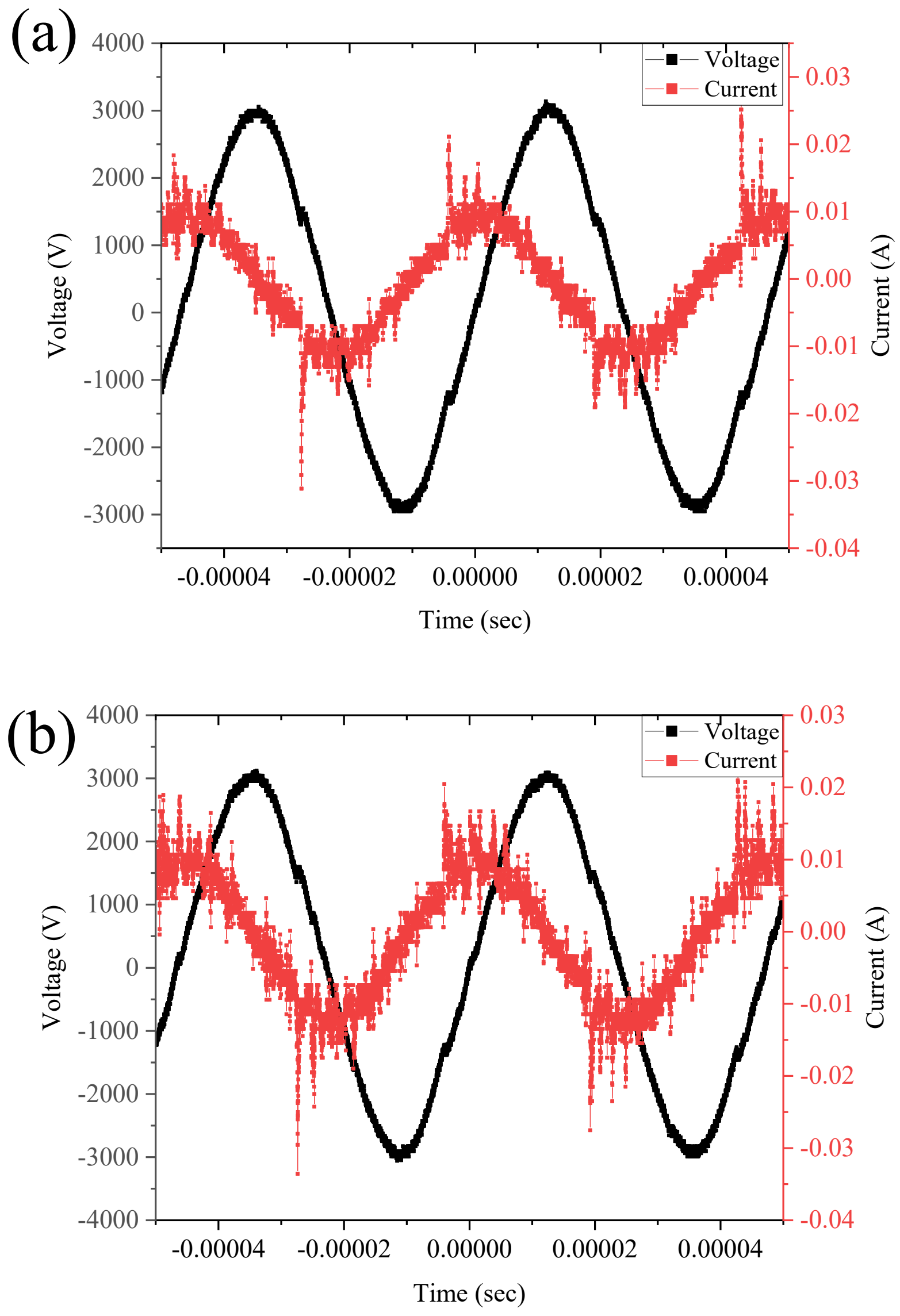

Figure 11. Voltage and current waveform, (a) rear inlet waveform, (b) side inlet waveform

Voltage and current waveforms (Figure. 11) were measured at 6.2 kV, 21.5 kHz, and 20 slm over 30 min with 5-min intervals. As shown in Figure 12, voltage remained stable throughout operation, with minimum values of 5.5 kV (rear) and 5.6 kV (side), supporting long-term device stability.

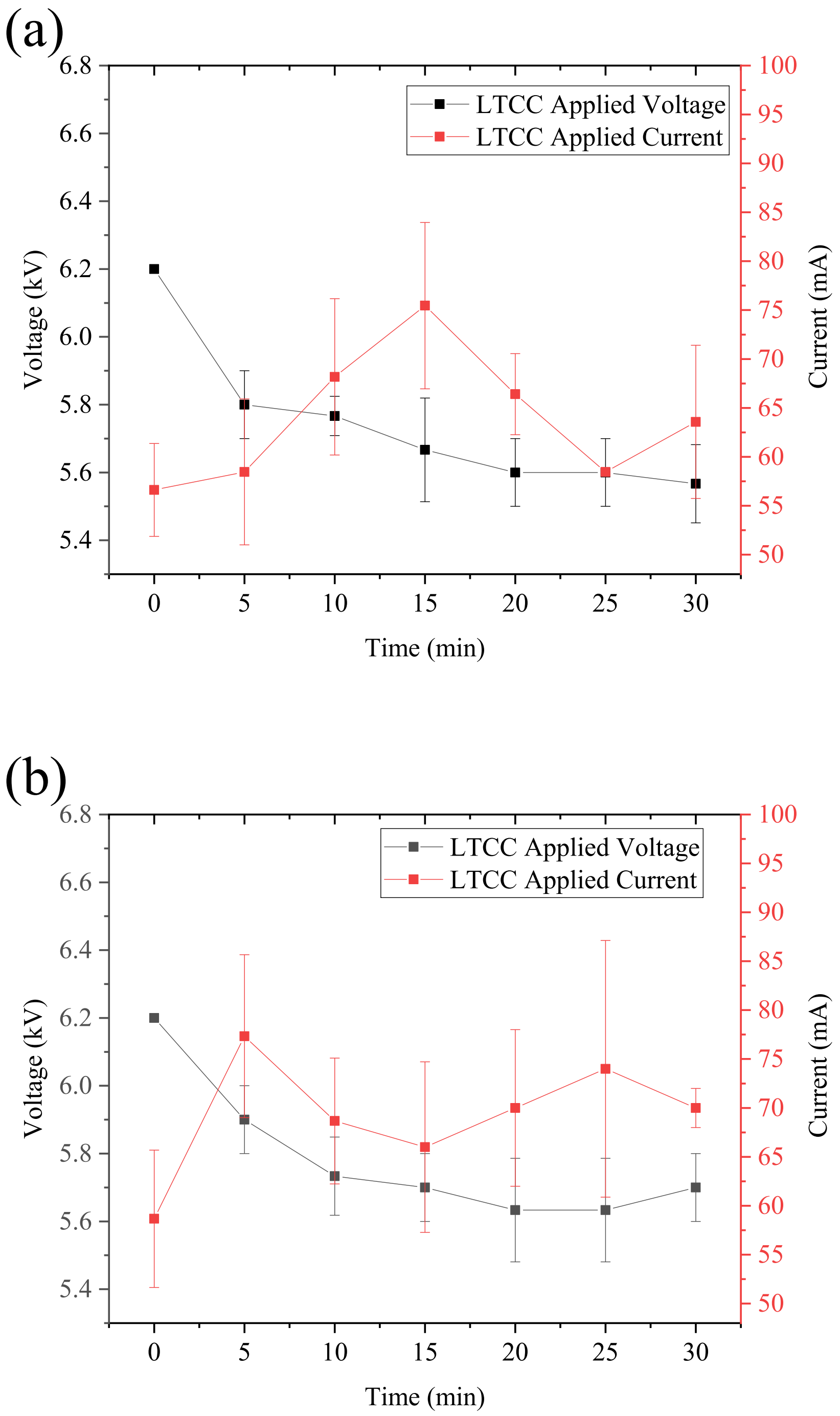

Figure 12. Voltage stability for a period of 30 minutes, (a) rear inlet setup, (b) side inlet setup

### 3.3.2 Lissajous Figures and Power Consumption Stability

Average power was 3.02 W (rear-inlet) and 3.07 W (side-inlet), consistent with higher side-inlet emission intensity. Figure 13 summarizes the time progression of total power. Power remained stable throughout the discharge for both setups, with the side-inlet consistently higher than the rear-inlet with P-value (P = 0.002) indicating a significant difference between the two configurations. The Lissajous plots revealed a noticeably larger vertical slope (dQ/dV) for the side-inlet case compared to the rear-inlet, reflecting greater charge transfer per voltage cycle during the discharge. This increased the charge transfer aligns with the side-inlet's higher emission intensity and greater degree of ionization.

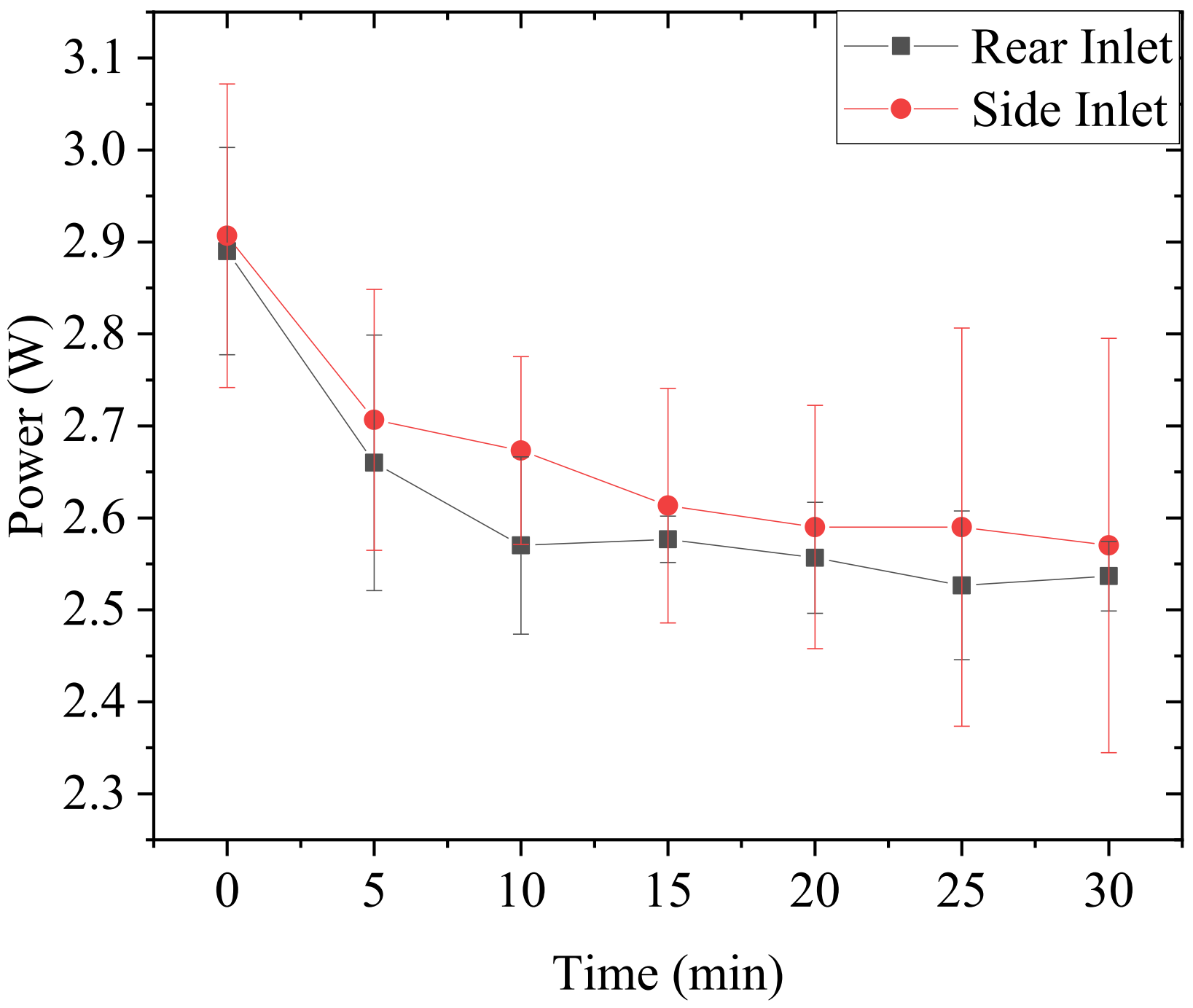

Figure 13. Power consumption stability

### 3.4 FLIR Temperature Measurements

#### 3.4.1 LTCC PPJ Surface Temperature

As shown in Figure 14 the device temperature stabilized after 5 min, reaching a maximum surface temperature of 48.4 °C (rear) and 50.5 °C (side), with the rear-inlet remaining cooler due to higher outlet velocity therefore enhancing its heat dissipation ($P=1.30\times10^{-4}$). Additionally, both configurations-maintained temperatures ≤ 50 °C, meeting the study's thermal criterion.

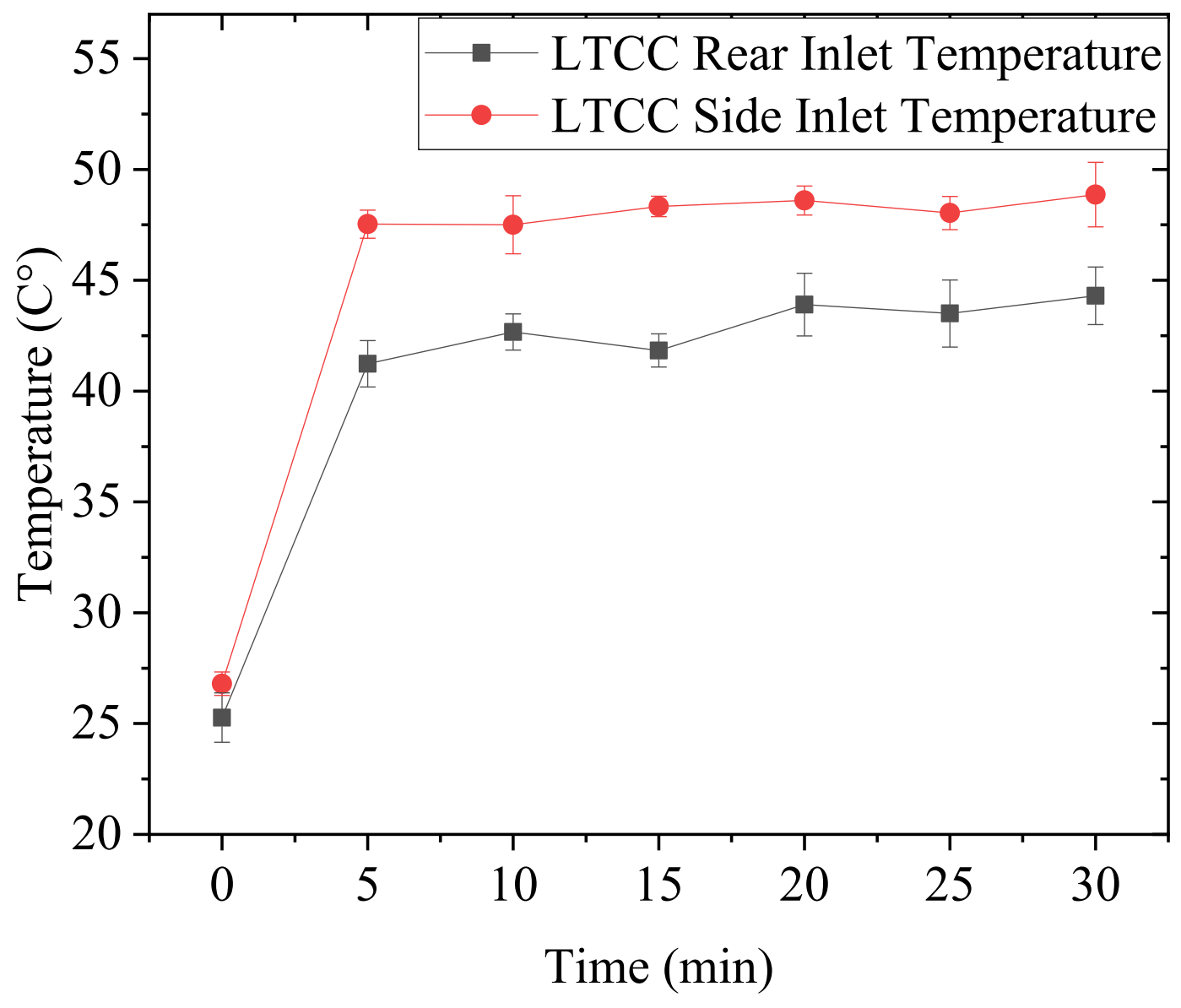

Figure 14. Temperature measurements for rear and side inlet configurations of PPJ surface

### 3.4.2 LTCC PPJ Gas Temperature

Figure 15 presents the gas temperature measured at the porcine-skin surface. A 5-min exposure was selected to approximate typical biomedical exposure times. Gas temperature measured at porcine skin reached ≤35 °C for both configurations, with no significant difference ($P = 0.67$). Low thermal conductivity of the porcine skin (0.2–0.3 W/m·K) as well as humidity found within the skin may have influenced the observed temperatures. Overall, the measured temperatures satisfy this study's thermal criterion and remain ≤40°C.

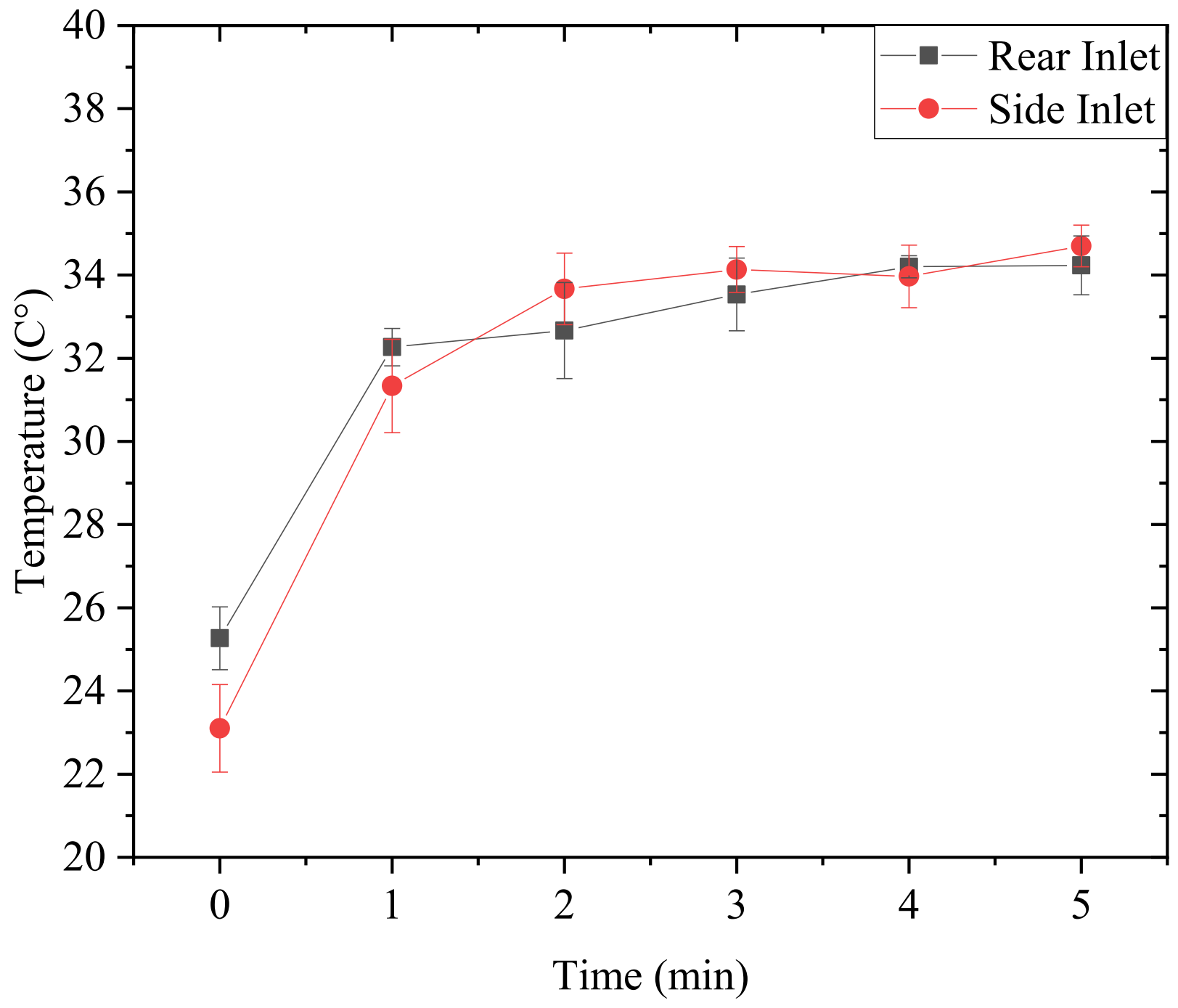

Figure 15. Gas temperature measurements for rear and side inlet configurations

### 3.5 Safety Parameters and Measurements

#### 3.5.1 UV Emission Intensity

The UV spectra for both inlet configurations showed a prominent emission band near 309 nm, attributed to the OH radical transition. The UV irradiance, weighted according to ICNIRP guidelines, was 0.64 W/m² for the rear-inlet and 2.20 W/m² for the side-inlet. The side-inlet configuration exhibited a significantly higher OH peak intensity than the rear-inlet, primarily due to enhanced ambient humidity promoting OH emission at this wavelength.

#### 3.5.2 Ozone Concentration

Ozone concentration produced by the rear- and side-inlet PPJ configurations was measured from 0–20 cm at 0° (on-axis) and 45° orientations. For the rear-inlet configuration, the on-axis concentration is low at 0 cm due to immediate rapid dispersion of gas when hitting the surface plate, ozone increased to a maximum of 0.531 ppm at 1 cm, and then decreased to 0.067 ppm at 20 cm. At 45°, the highest concentration occurred at 0 cm (0.623 ppm) and gradually decreased to 0.080 ppm at 20 cm.

The side-inlet configuration shows similar behavior but with slightly higher concentrations, reaching peaks of 0.682 ppm (0°) and 0.714 ppm (45°), and decreasing to 0.139 ppm and 0.128 ppm at 20 cm. This increase is consistent with stronger gas ionization in the side-inlet configuration, which promotes ozone formation.

These results indicate a recommended working distance of ≥20 cm. At this distance, rear-inlet ozone levels remain below established medical exposure limits (0.10–0.13 ppm), while side-inlet values fall within (45°) or slightly above (0°, 0.139 ppm) this range, suggesting that ventilation or flow conditioning may be required to maintain concentrations within recommended limits.

#### 3.5.3 Leakage Current

Patient leakage current was evaluated using an electrical human-body model consisting of $R_1 = 10\,\text{k}\Omega$, $R_2 = 1\,\text{k}\Omega$, and $C_1 = 0.015\,\mu\text{F}$. Plasma discharge was applied directly to an aluminum plate representing the body model, and leakage current was measured at working distances from 0 to 10 mm for both rear- and side-inlet configurations.

For both configurations, the measured leakage current decreases with increasing working distance. The current rose to approximately 130 μA at 1 mm and then gradually decreased with distance, reaching about 100 μA at 10 mm. Because the LTCC PPJ operates at 6.2 kV below the 14–15 kV required for stable plume formation the measured current primarily results from the electric field of the high-voltage electrodes. Therefore, the measured current primarily rose from the electric field of the high-voltage electrodes, which capacitively coupled to the aluminum plate of the human body model; increasing the separation reduced the coupling, and the measured current dropped.

No significant difference was observed between the rear- and side-inlet configurations ($P = 0.396$), indicating comparable electrical behavior. At a working distance of 10 mm, the leakage current decreases to approximately 100 µA, meeting the study's safety criterion (≤100 µA). Increasing the working distance further would be expected to reduce the leakage current proportionally.

# 4. Discussion

This work demonstrated a planar plasma jet (PPJ) realized in low-temperature co-fired ceramics (LTCC) with hermetically encapsulated high-voltage electrodes. Two gas-adapters, a rear-inlet and a side-inlet were implemented using a 9-channel adapter. Because a plasma plume in this design forms at 14–15 kV, beyond the preferred biomedical window, experiments were conducted at 6.2 kV and emphasized in-device characterization of discharge behavior, flow, and safety.

Outlet-velocity measurements showed that the rear-inlet produced higher gas speeds across the slit, whereas OES and ICCD indicated stronger emissions in the side-inlet channels. This difference is consistent with residence-time effects: higher velocity shortens excitation time and enhances convective cooling, while the lower velocity near the side-inlet sustains greater local ionization. Thermal data follow the same pattern peak surface temperatures of 48.4 °C (rear) and 50.5 °C (side) and effluent gas temperatures ≤ 35 °C for both indicating that the rear-inlet runs cooler overall. Electrical stability was confirmed by voltage–current waveforms and Lissajous analysis, with steady average powers of 3.02 W (rear-inlet) and 3.07 W (side-inlet); the modestly higher power for the side-inlet aligns with its greater emission and charge transfer.

The UV irradiance was 0.64 $W/m^2$ for the rear-inlet and 2.20 $W/m^2$ for the side-inlet, with the higher side-inlet intensity primarily driven by OH radical emission enhanced by ambient humidity. Ozone measurements at 0° and 45° decreased with stand-off, supporting a working distance ≥ 20 cm; at 20 cm the rear-inlet measured 0.067 ppm (0°) and 0.080 ppm (45°), while the side-inlet measured 0.139 ppm (0°) and 0.128 ppm (45°), values that can be driven fully into typical clinical ranges with modest dilution/ventilation or additional flow conditioning. Patient leakage current, quantified using an electrical human-body model, decreased with distance because the device does not form a plume at 6.2 kV and the coupling is primarily capacitive; at 10 mm the current reached 100 μA, meeting the ≤ 100 μA criterion, and larger stand-offs reduced it further. The side-inlet consistently exhibited slightly higher current, consistent with its stronger ionization and larger Lissajous charge transfer.

Trends predicted by simulations with higher outlet velocity for the rear-inlet and better gas distribution for the side-inlet were carried out experimentally, although channel-to-channel nonuniformity was greater than simulations suggested. The issue is attributed to manufacturing tolerances, surface roughness/porosity in printed structure, and non-ideal boundary conditions, and it points to specific design refinements: graded orifice area, refined channels, and smoother, non-porous flow material to equalize distribution.

Overall, the results show that LTCC encapsulation effectively stabilizes the discharge by isolating electrodes from ambient air, while gas-adapter architecture governs the trade-off between velocity-driven cooling/uniformity and ionization-driven intensity. Clinically acceptable operation will likely require coordinated control of flow uniformity, humidity linked OH emission (UV dose), ozone dilution, and varying distance to minimize capacitive coupling. These insights define clear data for the next design iteration toward a uniform, safe, and reproducible LTCC-based PPJ suitable for portable biomedical use.

## 5. Conclusion

This work demonstrated a compact planar plasma jet (PPJ) built in LTCC with hermetically encapsulated electrodes and two gas-delivery geometries (rear- and side-inlet) using a 9-channel adapter. Because a plasma plume requires 14–15 kV, characterization at 6.2 kV focused inside the device. Experiments confirmed stable operation and the expected flow–plasma trade-off: the rear-inlet produced higher outlet velocity and ran cooler, while the side-inlet showed stronger emission due to longer local residence time. Power was steady for both configurations (3.02 W-rear inlet and 3.07 W-side inlet), surface temperatures remained ≤50.5 °C, and gas temperatures ≤35 °C. Safety measurements highlighted environmental sensitivities: UV emissions at 309 nm were driven by high OH concentrations under humid conditions, ozone levels fell with distance and were acceptable at ≥20 cm (with one marginal side-inlet case), and patient leakage current dropped with stand-off to 100 μA at 10 mm. Flow uniformity was weaker than in simulation, pointing to practical effects of fabrication tolerances and surface roughness. Overall, LTCC encapsulation proved effective for electrical stability, while flow conditioning, humidity and UV control, and ozone mitigation emerged as critical design factors for safe PPJ operation.

## 6. Acknowledgement

The authors thanks National Science and Technology Council for the financial support.